\documentclass[runningheads]{llncs}
\usepackage[T1]{fontenc}
\usepackage{orcidlink}
\usepackage{hyperref}
\usepackage{marvosym}
\usepackage{textcomp}
\usepackage{amsmath}
\usepackage{algorithm}
\usepackage{algpseudocode}
\usepackage{amssymb}
\usepackage{graphicx}
\usepackage{bm} %
\begin{document}
\title{Simultaneous Tri-Modal Medical Image Fusion and Super-Resolution using Conditional Diffusion Model}
\titlerunning{Simultaneous Tri-Modal Medical Image Fusion and Super-Resolution}
%
\author{Yushen Xu\orcidlink{0009-0009-3931-8626}
        \and Xiaosong Li$^{(\textrm{\Letter})}$ \orcidlink{0000-0003-4672-1527}
        \and Yuchan Jie \orcidlink{0000-0001-9580-6012}
        \and Haishu Tan \orcidlink{0000-0001-8939-0452} } 

\authorrunning{Y. Xu et al}
%
\institute{Foshan University\\
\email{lixiaosong@buaa.edu.cn}}
\maketitle              
\begin{abstract}
In clinical practice, tri-modal medical image fusion, compared to the existing dual-modal technique, can provide a more comprehensive view of the lesions, aiding physicians in evaluating the disease's shape, location, and biological activity. However, due to the limitations of imaging equipment and considerations for patient safety, the quality of medical images is usually limited, leading to sub-optimal fusion performance, and affecting the depth of image analysis by the physician. Thus, there is an urgent need for a technology that can both enhance image resolution and integrate multi-modal information. Although current image processing methods can effectively address image fusion and super-resolution individually, solving both problems synchronously remains extremely challenging. In this paper, we propose TFS-Diff, a simultaneously realize tri-modal medical image fusion and super-resolution model. Specially, TFS-Diff is based on the diffusion model generation of a random iterative denoising process. We also develop a simple objective function and the proposed fusion super-resolution loss, effectively evaluates the uncertainty in the fusion and ensures the stability of the optimization process. And the channel attention module is proposed to effectively integrate key information from different modalities for clinical diagnosis, avoiding information loss caused by multiple image processing. Extensive experiments on public Harvard datasets show that TFS-Diff significantly surpass the existing state-of-the-art methods in both quantitative and visual evaluations. Code is available at \href{https://github.com/XylonXu01/TFS-Diff}{https://github.com/XylonXu01 /TFS-Diff}.

\keywords{Tri-Modal Medical Image Fusion  \and Super-Resolution \and Conditional Diffusion Model.}
\end{abstract}
\section{Introduction}
Multimodal medical images have become an indispensable tool in modern medical diagnosis and treatment planning. Computer Tomography (CT), Magnetic Resonance Imaging (MRI), Positron Emission Tomography (PET), and Single Photon Emission Computed Tomography (SPECT) each provide unique and complementary information \cite{1}, revealing the anatomical structure, physiological function, and molecular changes in the human body, respectively. However, due to the different imaging principles underlying these imaging technologies, images produced by various sensors exhibit significant differences in information content. Although the diversity of imaging enriches the sources of information for clinical diagnosis, it also poses additional challenges for physicians in integrating multi-source image information to make accurate diagnoses \cite{2}.

Multimodal image fusion holds promise for combining information from images of different modalities \cite{3,4} to obtain a more comprehensive diagnostic view. Currently, image fusion is mainly divided into methods based on deep learning \cite{6,7} and traditional methods \cite{9,10}. Deep learning-based methods often use generative adversarial networks (GANs) to simulate the distribution of fused images to obtain high-quality fused images \cite{13}. Although GAN-based methods can generate satisfactory fused images, they suffer from issues such as training instability, mode collapse and lack of interpretability. As an improvement, fusion methods based on Diffusion \cite{15,16} have been proposed, which generate high-quality images by simulating the diffusion process of restoring images corrupted by noise to clean images, thereby mitigating common problems like training instability and mode collapse in GANs, and their generation process is interpretable. For example, Zhao et al. \cite{17} proposed using DDPM for fusion tasks and employing a hierarchical Bayesian method to model the subproblems of maximum likelihood estimation. However, no deep learning fusion methods for tri-modal medical images have emerged, and only a few traditional methods have conducted preliminary research on this problem. For instance, Jie et al. \cite{19} proposed a tri-modal medical image fusion based on an adaptive energy selection scheme and sparse representation, using sparse representation to fuse texture components, and adaptive energy selection scheme to fuse cartoon components. Jie et al.\cite{20} proposed a tri-mode medical image fusion and denoising method based on BitonicX filtering. This method analyzes pixels in terms of gradient, energy, and sharpness to achieve medical image fusion and denoising.

Simultaneously, in medical imaging, due to various factors such as the resolution limits of imaging equipment, time constraints on image acquisition, and the radiation doses patients can tolerate, the resulting medical images often have limited resolution. Despite this, there is still an urgent demand for high-resolution (HR) medical images in clinical practice \cite{22}. Currently, deep learning methods for super-resolution can capture fine details and accurately preserve the original structure of images \cite{23,24,25}. For example, Mao et al. proposed a decoupled conditional diffusion model and extended it to multi-contrast MRI super-resolution, effectively estimating the uncertainty of the restoration and ensuring a stable optimization process \cite{11,28}. However, performing image super-resolution and image fusion in separate steps can propagate and amplify artifacts generated in the first step, thereby degrading the quality of the image. To address this issue, research has proposed end-to-end fusion and super-resolution methods for low-resolution images \cite{29,30,31}. For instance, Xiao et al. \cite{31} introduced a heterogeneous knowledge distillation network that embeds multi-layer attention to emphasize the texture details of visible light images and the prominent targets of infrared images to achieve both infrared and visible light image fusion and super-resolution simultaneously. However, this approach is only effective for the fusion of infrared and visible light images and lacks generalizability to medical images, and it cannot achieve fusion and super-resolution for three modalities.

Overall, although existing methods \cite{19,20,23} have achieved significant accomplishments in improving image quality, enhancing feature extraction, and improving single-modality medical image processing, they still face challenges in several key aspects: (1) Most research focuses on processing dual-modal source images, and there is a relative lack of in-depth studies on image processing problems that cover three modalities. (2) Current methods for medical image processing mostly focus on executing single tasks, lacking strategies for jointly optimizing fusion and super-resolution tasks. (3) Existing technologies that achieve both image fusion and super-resolution have limited generalization capabilities for tri-modal medical images.

To address these challenges, we propose an innovative Tri-modal Conditional Denoising Fusion-Super Resolution Diffusion model (TFS-Diff). To the best of our knowledge, this is the first study to achieve tri-modal medical image fusion and super-resolution tasks synchronously in an end-to-end manner. Our work's main contributions are threefold:
\begin{enumerate}
    \item The TFS-Diff model synchronously implements end-to-end tri-modal image fusion and super-resolution processing, eliminating the need for manually designing complex fusion and super-resolution network architectures. This significantly simplifies the model design process.
    \item We propose a feature fusion module based on a channel attention mechanism that can learn and extract shared features and modality-specific features from different modal medical images.
    \item A new fusion and super-resolution loss function is proposed to retain the sharpness, texture and contrast information of the medical images into the fused result. Meanwhile, it guarantees the stability of the model training process and the high quality of the fused results.
\end{enumerate}
\section{Method}
\subsection{Overall Architecture}
\begin{figure}
\centering	
\includegraphics[width=1.0\linewidth]{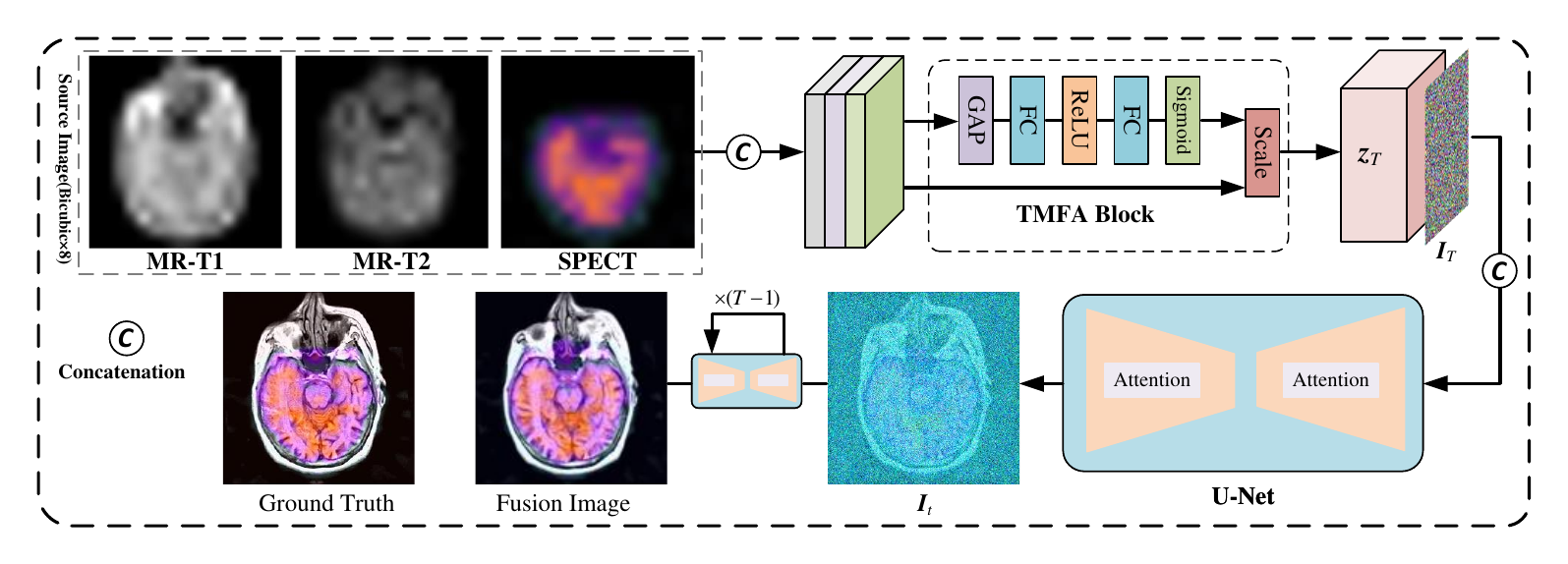}
\caption{Implementing super-resolution and fusion model structures synchronously. Taking MR-T1/MR-T2/SPECT fusion as an example, the original images need to be sampled to the specified resolution through bicubic interpolation as the input for the TFS-Diff model, and the output of TFS-Diff is compared with the Ground Truth to calculate the loss.}
\label{Fram}
\end{figure} 
For multimodal image fusion and super-resolution tasks, we propose TFS-Diff, a method based on a conditional denoising diffusion probability model. This method aims to generate high-resolution fused images that contain rich multimodal information and are highly consistent with the source images. As shown in Fig.~\ref{Fram}, TFS-Diff achieves precise tri-modal medical image fusion and its super-resolution through a forward and reverse Markov chain process.

Taking the fusion of MR-T1, MR-T2, and SPECT as an example, let the low-resolution of the three images be denoted as \(x \in R^{HW}\), \(y \in R^{HW}\), and \(s \in R^{HW}\) respectively, and the high-resolution fusion result be represented as \(\bm{I}_0 \in R^{3HW}\). The three modal images \(x\), \(y\), \(s\) are input into the model simultaneously, first upsampled to the target resolution through bicubic interpolation sampling of \(x\), \(y\), \(s\), and then feature extraction is performed through the TMFA Block (see Section 2.2) to obtain \(z_t = \varepsilon(x, y, s)\), which is concatenated on the channel dimension with \(\bm{I}_t \sim \mathcal{N}(\mathbf{0}, \bm{I})\). The objective function optimized by TFS-Diff is:
\begin{equation}
L_{\text{TFS}} := {E}_{\varepsilon{(x,y,s), \epsilon\sim \mathcal{N}(0,I),t }} \left[ \left\| \epsilon - \epsilon_\theta (z_t, I_t, \gamma_t) \right\|_2^2 \right]
\end{equation}
where, \(\gamma_t\) represents the noise variance, and \(\epsilon_{\theta}\) is the UNet \cite{32} used for noise prediction. \(z_t\) is obtained through the TMFA Block.

The backbone of TFS-Diff adopts the U-Net structure from SR3 \cite{33}, with \(z\) and \(t\) as inputs to \(\epsilon_{\theta}\). The backbone consists of a contracting path, an expansive path, and a diffusion head. Unlike the U-Net in DDPM \cite{16}, TFS-Diff uses residual blocks from BigGAN \cite{brock2018large}as connections and incorporates a self-attention mechanism. Both the contracting and expansive paths are comprised of 4 convolutional layers. The diffusion head consists of a single convolutional layer, used for generating predicted noise\cite{18}. Parameters are initialized using the Kaiming initialization method\cite{34}.

\subsection{Tri-modal Fusion Attention (TMFA) Block}
During the process of tri-modal medical image fusion, different modalities provide unique perspectives on anatomical structures, physiological functions, and molecular levels. Existing fusiong methods overlook the complementarity between modalities and the differences in feature importance across different channels. The main purpose of the TMFA Block is to extract deep feature of the concatenated multimodal images before entering the diffusion phase of the fusion super-resolution network. It utilizes a channel attention mechanism to learn the importance weights of each channel, enhancing useful features and suppressing irrelevant information. This approach highlights the most critical information for clinical diagnosis across different modalities.

As shown in Fig.\ref{Fram}, the structure of the TMFA Block is based on the classic SE (Squeeze-and-Excitation) block, which has been customized to adapt to tri-modal image. Initially, a Global Average Pooling (GAP) layer compresses the features of the input images \( C = \text{concatenate}(x, y, s) \) to capture global context information. Subsequently, a bottleneck structure composed of two Fully Connected (FC) layers is introduced to learn the nonlinear relationships between channels, where the ReLU activation function is applied to the first FC layer, and the Sigmoid activation function is applied to the second FC layer, to output the attention weights of the channels. Finally, these attention weights are utilized to adjust the importance of each channel in the original feature image, thus accomplishing feature recalibration to obtain the feature \(Z_c\).
\subsection{Fusion super-resolution joint loss function}
To make TFS-Diff training converge more stably, a new joint loss design is implemented\(L_{PSF}\), combining Mean Squared Error (MSE) loss and Structural Similarity Index (SSIM) loss. \(L_{PSF}\) leverages the advantages of MSE loss in terms of pixel-level reconstruction accuracy, as well as the effectiveness of SSIM loss in maintaining image structural information and enhancing visual quality to optimize both the pixel accuracy and visual quality of the image simultaneously.
\begin{gather}
    L_{PSF}={\lambda_1}{L}_{MSE}+{\lambda_2}{L}_{ssim}\\
  {L}_{MSE} = \frac{1}{N} \sum_{i=1}^{N} (P_i - T_i)^2\\
{L}_{SSIM}=\frac{\left(2\mu_P\mu_T+c_1\right)\left(2\sigma_{PT}+c_2\right)}{\left(\mu_P^2+\mu_T^2+c_1\right)\left(\sigma_P^2+\sigma_T^2+c_2\right)}
\end{gather}
where \(\lambda_1, \lambda_2 \in (0,1]\) represent the weights of the two losses, respectively.
\section{Experiments}
\subsection{Experimental Detail}
The dataset covers five different types of registered medical images, including MR-T2/MR-Gad/PET, CT/MR-T2/SPECT, MR-T1/MR-T2/PET, MR-T2/MR-Gad/SPECT, and MR-T1/MR-T2/SPECT. All source images are from the whole brain atlas database of Harvard Medical School\cite{summers2003harvard}, We randomly divided the data into 84, 10 and 25 groups as training set, validation set and test set respectively.The resolution of the training and testing images is 256x256, which was downsampled using bicubic interpolation to construct super-resolution datasets with different magnification levels (8x, 4x, 2x).The Ground Truth is fused by the BitonicX \cite{20}.

Five state-of-the-art (sota) methods were used for comparison, including three dual-modal fusion methods: CDDFuse \cite{7}, TGFuse \cite{35}, DDFM \cite{17}, and two tri-modal fusion methods: BitonicX Filtering \cite{20}, CTSR \cite{19}. Additionally, the SR3 model \cite{33} was used as the baseline for super-resolution.

The model was optimized using the Adam optimizer with a fixed learning rate of 1e-4 and a diffusion step count \(T\) of 4000. The model was trained for 800,000 steps on a computer equipped with four NVIDIA GeForce RTX 3090 GPUs, with a batch size set to 32.

\subsection{Objective evaluation metric}
In this study, we evaluated the proposed model's performance using several quantitative metrics: Average Gradient (AG) \cite{AG}, Mean Squared Error (MSE) \cite{MSE}, Visual Information Fidelity (VIF) \cite{VIF}, Structural Similarity Index (SSIM) \cite{PSNR}, Peak Signal-to-Noise Ratio (PSNR)\cite{PSNR}, Perceptual Image Quality Loss (LPIPS) \cite{LPIPS}, Mean Absolute Error (MAE), and Root Mean Squared Error (RMSE). These metrics assess the model's performance in various aspects: AG measures image edge and texture clarity; MSE, MAE, and RMSE gauge pixel-level accuracy; VIF evaluates visual quality; SSIM and PSNR judge structural similarity and noise ratio, indicating image visual effects and fidelity; LPIPS assesses perceptual quality from a deep learning perspective. 
\subsection{Comparison with SOTA methods}
\begin{figure}
\centering	
\includegraphics[width=1\linewidth]{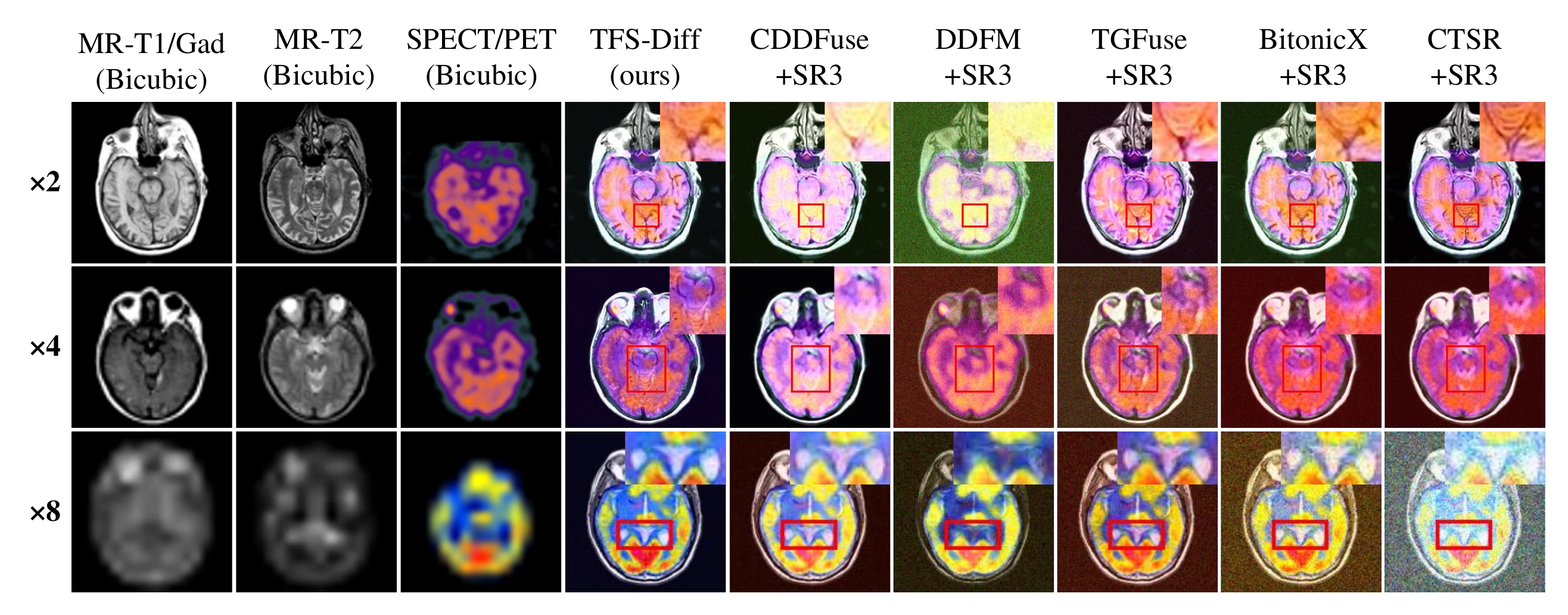}
\caption{Tri-modal fusion on the Harvard dataset showed super-resolution results at amplification factors 2, 4, and 8, using MR-T1/MR-T2/SPECT, MR-Gad/MR-T2/SPECT, and MR-T1/MR-T2/PET configurations, respectively.}
\label{Compare}
\end{figure} 
Table \ref{table1} and Fig.\ref{Compare} respectively show the objective evaluation metrics and fusion results of TFS-Diff, CDDFuse, DDFM, TGFuse, BitonicX and CTSR methods. From Table \ref{table1}, it can be observed that TFS-Diff ranks first in all evaluation metrics, demonstrating its outstanding performance. Specifically, under the conditions of resolutions enlarged by ×8, ×4, and ×2, the methods of TGFuse, DDFM and CDDFuse rank second to TFS-Diff in overall objective evaluation metrics, respectively. Additionally, Fig.\ref{Compare} clearly reveals the problem that the SR3 model, when applied to the DDFM method for super-resolution, fails to completely eliminate noise at three different magnification rates. As the magnification rate increases, the quality of the fused images obtained by the CTSR, TGFuse, and BitonicX methods declines. In contrast, TFS-Diff shows its effectiveness in maintaining the texture and color information of the source images under the three types of magnifications.Due to space constraints, more detailed renderings will be shown in the supplemental documentation

\begin{table}[]
\centering
\caption{The objective results for Harvard dataset (Bold: the best; comparison methods all use SR3 for super-resolution)}
\label{table1}
\begin{tabular}{lllllllllllllll}
\cline{1-9}
Scale &
  \multicolumn{8}{c}{×2} &
   &
   &
   &
   &
   &
   \\
Metrics &
  MSE $\downarrow$ &
  VIF $\uparrow$ &
  SSIM $\uparrow$ &
  PSNR $\uparrow$ &
  LPIPS $\downarrow$ &
  MAE $\downarrow$ &
  RMSE $\downarrow$ &
  AG $\uparrow$ &
   &
   &
   &
   &
   &
   \\ \cline{1-9}
BitonicX &
  2634.907 &
  0.471 &
  0.579 &
  14.42 &
  0.419 &
  53.95 &
  110.82 &
  8.913 &
   &
   &
   &
   &
   &
   \\
CDDFuse &
  3519.701 &
  0.501 &
  0.759 &
  12.74 &
  0.339 &
  54.92 &
  120.62 &
  8.237 &
   &
   &
   &
   &
   &
   \\
CTSR &
  2733.143 &
  0.452 &
  0.611 &
  14.11 &
  0.435 &
  55.74 &
  114.17 &
  9.436 &
   &
   &
   &
   &
   &
   \\
DDFM &
  2920.907 &
  0.480 &
  0.549 &
  13.50 &
  0.496 &
  56.96 &
  107.37 &
  8.237 &
   &
   &
   &
   &
   &
   \\
TGFuse &
  3013.091 &
  0.479 &
  0.600 &
  14.13 &
  0.424 &
  54.58 &
  110.61 &
  9.945 &
   &
   &
   &
   &
   &
   \\
\textbf{Ours} &
  \textbf{2021.23} &
  \textbf{0.577} &
  \textbf{0.818} &
  \textbf{15.27} &
  \textbf{0.319} &
  \textbf{44.80} &
  \textbf{103.64} &
  \textbf{9.959} &
   &
   &
   &
   &
   &
   \\ \cline{1-9}
Scale &
  \multicolumn{8}{c}{×4} &
   &
   &
   &
   &
   &
   \\
Metrics &
  MSE $\downarrow$ &
  VIF $\uparrow$ &
  SSIM $\uparrow$ &
  PSNR $\uparrow$ &
  LPIPS $\downarrow$ &
  MAE $\downarrow$ &
  RMSE $\downarrow$ &
  AG $\uparrow$ &
   &
   &
   &
   &
   &
   \\ \cline{1-9}
BitonicX &
  2010.889 &
  0.424 &
  0.743 &
  15.26 &
  0.432 &
  49.16 &
  106.78 &
  7.475 &
   &
   &
   &
   &
   &
   \\
CDDFuse &
  3390.033 &
  0.430 &
  0.671 &
  13.08 &
  0.415 &
  56.96 &
  119.82 &
  7.732 &
   &
   &
   &
   &
   &
   \\
CTSR &
  2159.267 &
  0.426 &
  0.672 &
  14.99 &
  0.439 &
  49.74 &
  107.16 &
  7.577 &
   &
   &
   &
   &
   &
   \\
DDFM &
  2511.116 &
  0.458 &
  0.694 &
  13.90 &
  0.489 &
  49.04 &
  99.45 &
  7.082 &
   &
   &
   &
   &
   &
   \\
TGFuse &
  2213.12 &
  0.448 &
  0.715 &
  15.32 &
  0.415 &
  47.86 &
  102.75 &
  8.109 &
   &
   &
   &
   &
   &
   \\
\textbf{Ours} &
  \textbf{1740.448} &
  \textbf{0.560} &
  \textbf{0.788} &
  \textbf{15.78} &
  \textbf{0.340} &
  \textbf{46.11} &
  \textbf{97.98} &
  \textbf{8.152} &
   &
   &
   &
   &
   &
   \\ \cline{1-9}
Scale &
  \multicolumn{8}{c}{×8} &
  \multicolumn{1}{c}{} &
  \multicolumn{1}{c}{} &
  \multicolumn{1}{c}{} &
  \multicolumn{1}{c}{} &
  \multicolumn{1}{c}{} &
  \multicolumn{1}{c}{} \\
Metrics &
  MSE $\downarrow$ &
  VIF $\uparrow$ &
  SSIM $\uparrow$ &
  PSNR $\uparrow$ &
  LPIPS $\downarrow$ &
  MAE $\downarrow$ &
  RMSE $\downarrow$ &
  AG $\uparrow$ &
   &
   &
   &
   &
   &
   \\ \cline{1-9}
BitonicX &
  4413.324 &
  0.394 &
  0.532 &
  12.14 &
  0.521 &
  64.89 &
  124.32 &
  13.558 &
   &
   &
   &
   &
   &
   \\
CDDFuse &
  5310.433 &
  0.378 &
  0.511 &
  11.21 &
  0.535 &
  70.20 &
  132.68 &
  12.671 &
   &
   &
   &
   &
   &
   \\
CTSR &
  5480.484 &
  0.369 &
  0.568 &
  11.56 &
  0.549 &
  72.41 &
  135.18 &
  13.612 &
   &
   &
   &
   &
   &
   \\
DDFM &
  4998.928 &
  0.307 &
  0.466 &
  11.50 &
  0.622 &
  68.10 &
  125.33 &
  12.337 &
   &
   &
   &
   &
   &
   \\
TGFuse &
  4142.136 &
  0.401 &
  0.566 &
  12.69 &
  0.508 &
  62.77 &
  121.60 &
  13.193 &
   &
   &
   &
   &
   &
   \\
\textbf{Ours} &
  \textbf{1559.803} &
  \textbf{0.579} &
  \textbf{0.980} &
  \textbf{16.50} &
  \textbf{0.299} &
  \textbf{40.68} &
  \textbf{96.42} &
  \textbf{13.77} &
   &
   &
   &
   &
   &
   \\ \cline{1-9}
\end{tabular}

\end{table}

\subsection{Ablation Study}
This section aims to verify the effectiveness and contribution of the TMFA Block and PSF Loss in our TFS-Diff.
\begin{enumerate}
    \item \textbf{w/o TMFA Block}: We removed the TMFA Block to assess its significance in feature extraction and information fusion processes.
    \item \textbf{w/o PSF Loss}: We replaced PSF Loss with MSE loss to evaluate the importance of PSF Loss in balancing pixel accuracy and visual quality.
\end{enumerate}
\begin{table}[]
\centering
\caption{Ablation Study on the Harvard dataset(Bold: the best).}
\label{Ablation}
\begin{tabular}{lllllllll}
\hline
\multicolumn{9}{c}{Harvard dataset(×2)}                                             \\
Metrics        & VIF$\uparrow$  & SSIM$\uparrow$   & PSNR$\uparrow$  & AG$\uparrow$    & MSE$\downarrow$     & LPIPS$\downarrow$  & MAE$\downarrow$   & RMSE$\downarrow$   \\ \hline
w/o TMFA Block & 0.468 & 0.761  & 14.63 & 9.529 & 2777.63 & 0.362  & 60.60 & 120.18 \\
w/o PSF Loss   & 0.452 & 0.802  & 14.76 & 9.021 & 2171.51 & 0.351  & 55.07 & 120.81 \\
TFS-Diff       & \textbf{0.577} & \textbf{0.8180} & \textbf{15.27} & \textbf{9.959} & \textbf{2021.23} & \textbf{0.319} & \textbf{44.80} & \textbf{103.64} \\ \hline
\multicolumn{9}{c}{Harvard dataset(×4)}                                             \\
Metrics        & VIF$\uparrow$  & SSIM$\uparrow$  & PSNR$\uparrow$ & AG$\uparrow$   & MSE$\downarrow$    & LPIPS$\downarrow$ & MAE$\downarrow$  & RMSE$\downarrow$  \\ \hline
w/o TMFA Block & 0.558 & 0.746  & 15.32 & 7.13 & 1907.22 & 0.404  & 56.42 & 114.96 \\
w/o PSF Loss   & 0.545 & 0.728  & 15.67 & 7.801 & 1761.69 & 0.365  & 51.42 & 112.71 \\
TFS-Diff       & \textbf{0.560} & \textbf{0.788} & \textbf{15.78} & \textbf{8.152} & \textbf{1740.44} & \textbf{0.340} & \textbf{46.10} & \textbf{97.98}  \\ \hline
\multicolumn{9}{c}{Harvard dataset(×8)}                                             \\
Metrics        & VIF$\uparrow$  & SSIM$\uparrow$  & PSNR$\uparrow$  & AG$\uparrow$   & MSE$\downarrow$    & LPIPS$\downarrow$ & MAE$\downarrow$  & RMSE$\downarrow$  \\ \hline
w/o TMFA Block & 0.539 & 0.792  & 15.40 & 13.60 & 1871.20 & 0.347  & 45.81 & 105.15 \\
w/o PSF Loss   & 0.550 & 0.569  & 15.63 & 12.67 & 1712.41 & 0.354  & 50.82 & 108.77 \\
TFS-Diff       & \textbf{0.579} & \textbf{0.980} & \textbf{16.50} & \textbf{13.77} & \textbf{1559.80} & \textbf{0.299} & \textbf{40.68} & \textbf{96.42}  \\ \hline
\end{tabular}
\end{table}
As shown in the Table \ref{Ablation}, the ablation study results confirmed the significant contribution of the TMFA Block and PSF Loss to enhance the performance of the tri-modal medical image fusion super-resolution model. The combination of these components not only optimized pixel-level reconstruction but also significantly improved the visual quality of the images, thus providing an effective solution for complex medical image processing tasks.
\section{Conclusion}
This study proposed a conditional diffusion model, TFS-Diff, for tri-modal medical image fusion super-resolution, introducing two key innovations: the TMFA Block and PSF Loss, which ensure the generation accuracy of the diffusion model. Through comprehensive experimental validation, our approach has achieved significant improvements in detail restoration and visual quality compared to existing technologies.

The TMFA block optimized the model's capability to fuse features from different modal medical images, enhancing the efficiency and quality of information integration. Simultaneously, the design of PSF Loss successfully balanced pixel-level accuracy and structural similarity, further enhancing the model's performance in image reconstruction. Ablation results confirmed the significant contribution of these two components to model performance improvement, reflecting their application value in complex medical image processing tasks. Moreover, if TFD-Diff is applied to tri-modal medical image fusion medical instruments, it will enhance the diagnostic efficiency of doctors and reduce the time and money spent by patients.

Our future work will integrate large language models to assist in modeling the diffusion process of TFS-Diff, which is expected to enhance the model's generalizability across various types of medical images.


\begin{credits}
\subsubsection{\ackname} This research was supported in part by the National Natural Science Foundation of China under Grant 62201149, in part by the Basic and Applied Basic Research of Guangdong Province under Grant 2023A1515140077, in part by the Natural Science Foundation of Guangdong Province under Grant 2024A1515011880, in part by the Guangdong Higher Education Innovation and Strengthening of Universities Project under Grant 2023KTSCX127, and in part by the Foshan Key Areas of Scientific and Technological Research Project under Grant  2120001008558, China.

\subsubsection{\discintname}
The authors have no competing interests to declare that are
relevant to the content of this article.
\end{credits}

%
%
%
\bibliographystyle{splncs04}
\bibliography{Paper-3901}

%





\end{document}